# Photoactivation of neurons by laser-generated local heating

Benjamin Migliori[*] (1), Massimiliano Di Ventra (1), and William Kristan Jr (2)

(1) Department of Physics, University of California, San Diego
(2) Department of Neuroscience, University of California, San Diego

We present a method for achieving temporally and spatially precise photoactivation of neurons without the need for genetic expression of photosensitive proteins. Our method depends upon conduction of thermal energy via absorption by a dye or carbon particles and does not require the presence of voltage-gated channels to create transmembrane currents. We demonstrate photothermal initiation of action potentials in *Hirudo verbana* neurons and of transmembrane currents in *Xenopus* oocytes. Thermal energy is delivered by focused 50 ms, 650 nm laser pulses with total pulse energies between 250 and 3500 µJ. We document an optical delivery system for targeting specific neurons that can be expanded for multiple target sites. Our method achieves photoactivation reliably (70 - 90% of attempts) and can issue multiple pulses (6-9) with minimal changes to cellular properties as measured by intracellular recording. Direct photoactivation presents a significant step towards all-optical analysis of neural circuits in animals such as *Hirudo verbana* where genetic expression of photosensitive compounds is not feasible.

---

[*] Author to whom correspondence should be addressed.  Electronic mail: migliori@ucsd.edu

**I. Introduction**

The use of genetically expressed photo-sensitive channels in neurons has dramatically expanded the ability to generate spatially and temporally precise stimuli for probing neural circuit functions[1]. Traditional methods require the use of stimulating electrodes that are manually fixed in a static position during an experiment. By replacing electrodes with optical methods, stimulation patterns can be dynamically changed[2] or performed in awake behaving animals[3]. In the expanding body of research using photoactivation of neurons, most methods rely on genetic expression of specific channels[1,4]. However, in any of the animal models such as *Hirudo* where genetic engineering is difficult (see [5]), an alternative method would be very useful for investigating the remarkable and complex behaviors found at the circuit level[6,7]. Voltage-sensitive dyes now allow recording from many leech neurons simultaneously, revealing behavioral activity patterns[8]. As advances in optical recording of leech neuronal potentials increase the temporal and spatial resolution of optical electrophysiology, the need for a method of photoactivating neurons to affect neuronal activity patterns has become more crucial. Such a method would allow investigators to initiate "fictive" behaviors (in which motor patterns are exhibited in the isolated nervous system[9]) via dynamic and physiologically relevant stimuli.

Earlier experiments of Farber and Grinvald[10] achieved photoactivation without genetic expression using potentiometric dyes such as Neurodye RGA-30. These authors suggested small nonspecific transient[11] pores formed by optical dye breakdown as the mechanism underlying RGA-30 stain-mediated photosensitization, but the method was not widely adopted and the mechanism for activating the neurons was not fully clarified.

Our aim was to optimize and characterize photoactivation by optical absorbers including RGA-30. We reasoned that, if heat were the underlying cause of the membrane potential changes, then chemically neutral broad-band absorbers such as powdered carbon should also be capable of inducing action potentials when irradiated with a highly focused laser. These neutral absorbers are preferable to dyes because they do not interfere with the electrophysiological properties of leech neurons or with recordings using potentiometric dyes. Specifically, we investigated whether temperature-dependent conductance changes in biological membranes or thermal activation of voltage-gated channels (VGCs) could induce observed changes in membrane potential during laser irradiation. We also demonstrated a cost effective experimental design capable of direct laser photoactivation. We accomplished this by implementing an optical relay system based on multiple-trap optical tweezers powered by a low-cost laser diode[12].

II. **Materials and Methods**

**A. Preparation**

Photoactivation experiments were carried out in isolated adult *Hirudo* segmental ganglia and in stage VI *Xenopus* oocytes. Leeches were obtained from Niagara Leeches (Niagara Falls, NY) or our own breeding colony. Segmental ganglia were prepared for recording as described previously[13]. Wild-type Stage VI oocytes were harvested from surgically removed ovaries (Nasco, WI). We used standard procedures to prepare oocytes for recording[14,15].

**B. Desheathing and staining**

Segmental ganglia in *Hirudo* consist of approximately 400 unipolar cell bodies arranged around an inner neuropil and protected by a transparent outer sheath[16]. To stain neurons or deposit carbon particles on their somata, this sheath was removed using a micro-surgical blade (10316-14 FST). Properly desheathed ganglia are physiologically indistinguishable from sheathed preparations.

Desheathed ganglia were bathed for 20 minutes with 20mM Neurodye RGA-30 (Cole-Parmer) in cooled (approximately 5º C) leech saline flowing through a peristaltic pump[8]. After 20 minutes, we switched to a solution of leech saline without dye and flushed the ganglion for an additional 5 minutes.

**C. Electrophysiology**

Prepared oocytes or ganglia were placed on a translating sample stage consisting of a dish, micropipette manipulators, and a darkfield illumination system (Figure 1). Impalement of cells was visualized with a dissecting microscope. Leech neuronal membrane potentials were recorded intracellularly from individual cell bodies using 30–40 MΩ micropipettes filled with 3 M KAc; the signal was measured with an Axoclamp 2B amplifier (Axon Instruments, Sunnyvale, CA). Oocyte membrane potentials were recorded with 3 M KCl-filled 1-3 MΩ micropipettes and the signal measured by an Axon Instruments GeneClamp 500B amplifier. Current was injected through the micropipettes to induce depolarizations or action potentials. We used the WinWCP program for experimental control (Strathclyde University, freely available at http://spider.science.strath.ac.uk/sipbs/showPage.php?page=software_ses). After successful impalement, the sample stage was translated into the photoactivation optical pathway.

**D. Deposition of carbon particles**

Carbon sticks (Blick Art Supplies, San Diego, CA) were placed in a clean mortar and ground for approximately 20 minutes. Visual inspection of the resulting powder revealed particles ranging in size from 2 - 20 µm. Approximately 1 g of carbon powder was suspended in 3 ml of leech or oocyte saline at 5º C. After impaling a target cell, we pipetted 500 µL of the carbon-saline mixture onto the preparation, producing an ash-like deposition of carbon. We impaled cells before depositing the carbon powder because the layer of opaque particles made it difficult to successfully impale specific cells. We targeted isolated 2 - 10 µm particles for laser irradiation.

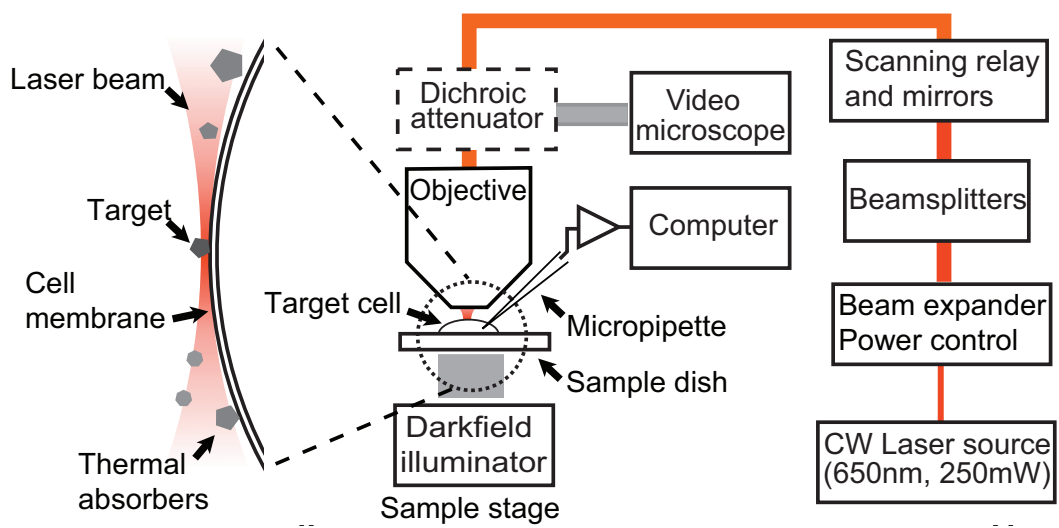

**Figure 1. Experiment Block Diagram**
Photoactivation is achieved by laser irradiation (pathway in red) of a target (either a carbon particle or a dye molecule) proximal to a neural cell membrane with sufficiently high absorption at a chosen wavelength. The target and neural preparation are visualized with a video darkfield system (pathway in grey). Cell responses are recorded with a computer using intracellular electrophysiological recordings. The elements of the laser pathway are discussed in the text.

### E. Photoactivation

We produced photoactivation using focused laser irradiation (Figure 1) from a 250mW, 650nm continuous wave diode laser (AiXiZ Lasers). A 3x beam expander was used to correct small divergence errors in the beam output[17]. The beam was split into polarization components by a half-wave plate and a polarizing beamsplitter, then recombined by a second polarizing beamsplitter. By rotating the half-wave plate to control the amount of power in each polarization mode and rejecting one of the pathways, we could smoothly vary the beam power. To change the beam target position without changing input power, we controlled the angle of the beams using gimbal-mounted mirrors in a scanning relay. The recombined and expanded beam image was relayed from these mirrors into the back aperture of a water immersion objective lens (Olympus XLUMPFL 20x, 0.95NA).

### F. Target alignment and visualization

We spatially targeted the focused laser by attenuating it to 0.1 mW with a removable IR longpass dichroic mirror mounted above the objective. With the target cell impaled and the IR dichroic in place, we opened the laser shutter and focused the laser onto a thermal absorber (powdered carbon or dye) in contact with the cell membrane (Figure 1). We then closed the shutter and removed the attenuator to allow full power throughput. We controlled the timing of the laser pulses during stimulation using a shutter (Uniblitz) driven by the WinWCP program.

We visualized the laser focus and target position by darkfield illumination using a white LED source (Luxeon 3W) mounted on a 3-axis manipulator under the sample dish. The darkfield image was reflected by the IR dichroic and imaged using a CCD camera (C-mount, Hitachi).

### G. Power and temperature measurement

We measured the power of the laser spot after the objective using a LaserCheck (Edmund Scientific). To estimate the upper bound of temperature increase during a laser pulse, we exposed a carbon layer on a microscope slide, immersed in leech saline at 20º C, to 50 ms of laser irradiation at various energies. At typical maximum power (70mW, 3500 uJ) this consistently resulted in focal boiling of the liquid as evidenced by rapid bubble formation.

## III. Results

### A. Neutral photoabsorbers photosensitize neurons with high reliability

During our laser irradiation experiments, we estimated the upper bound of the temperature change during laser exposure to be 50 - 90º C (see methods). This is based on the finding that bubbles from focal boiling occur within milliseconds with irradiances of 23 mW/µm$^2$ (3500 µJ pulse energy), and we used 1.5 - 5 mW/µm$^2$ (250 - 750 µJ pulse energy) to activate neurons. Because we would not expect the relatively low energy photons of the 650nm laser to drive photochemical reactions, and the peak laser power is not high enough to induce optical breakdown, we hypothesize that the primary mechanism of photoactivation depends upon thermal energy conduction. Thermal energy is absorbed at 650 nm by many different compounds, but we chose to use powdered carbon because it is relatively chemically neutral. We found that irradiation of a carbon particle on the surface of a neuron with 50 ms, 250 - 700 µJ laser pulses induced action potentials in all types of neurons sampled, which included Retzius (Rz), AP, P, and T cells (Figure 2a[i]). The observed depolarizations are qualitatively similar to photoactivation achieved using RGA-30 dye in the AP and Rz cell membranes (Figure 2b[i]). The typical response was a single action potential during the 50ms window, although trains of action potentials were sometimes observed.

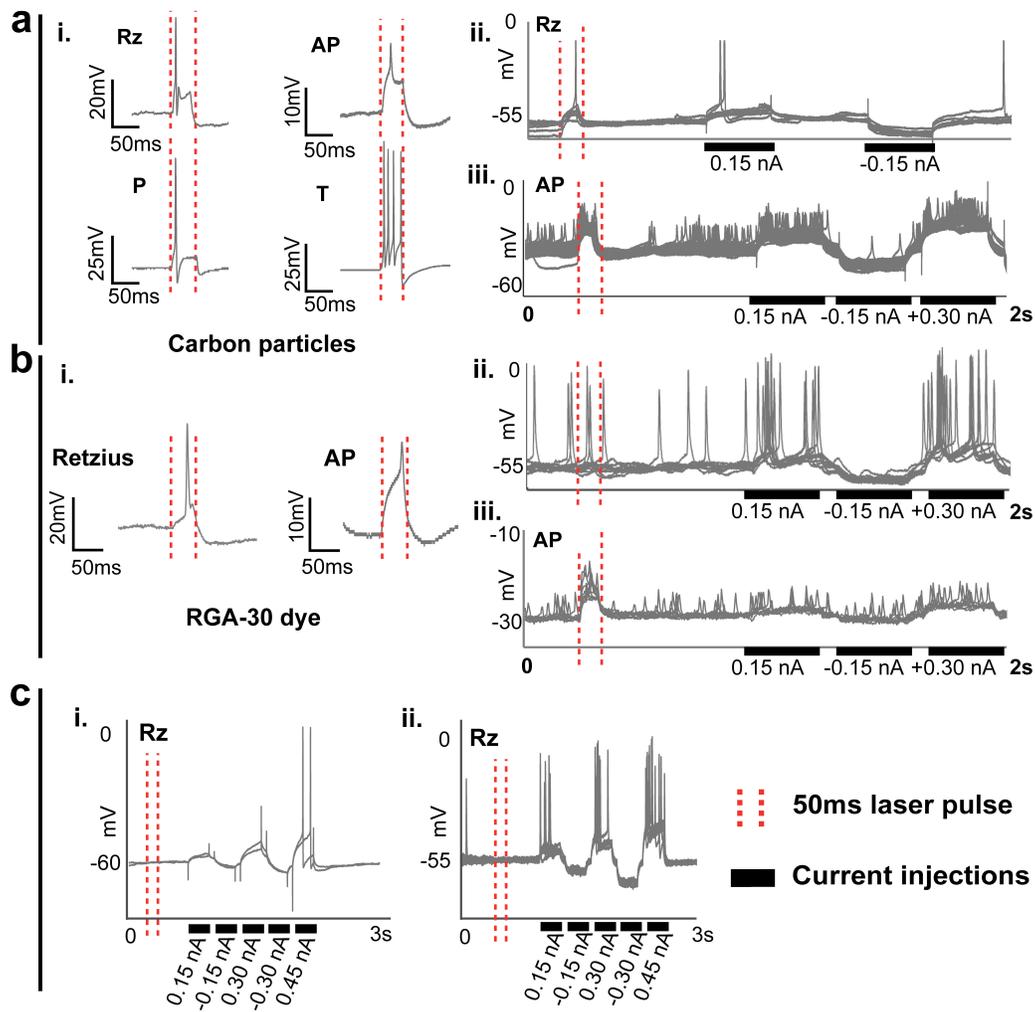

**Figure 2: Neural responses to laser irradiation**
(a)[i] Intracellular recordings from Retzius (Rz), AP, P, and T cells during laser irradiation of carbon particles with 50 ms, 250-700 µJ laser pulses (duration indicated by red dashed lines). (a)[ii] Intracellularly recorded Rz cell responses to repeated laser irradiation (250 µJ) of carbon particles. 10 responses shown superimposed. At this intensity, 10% of the laser pulses elicited action potentials, and 100% elicited depolarizations. (a)[iii] Intracellularly recorded AP cell responses to repeated laser irradiation (250 µJ) of carbon particles. 40 episodes shown superimposed; 90% of the pulses elicited action potentials and 100% elicited depolarization. Depolarizing and hyperpolarizing current pulses (labelled black bands) were injected during each episode in a[ii] and a[iii] to determine the electrophysiological state of each cell. (b)[i] Intracellular recordings from Retzius and AP cells stained with the dye RGA-30 during 50 ms, 700 µJ laser irradiation (duration indicated by red dashed lines). (b)[ii] Intracellularly recorded Retzius cell responses to repeated 700 µJ laser irradiation of the dye RGA-30. 10 responses superimposed; 20% of the pulses elicited action potentials and 60% elicited depolarizations. (b)[iii] Intracellularly recorded AP cell response to repeated 50 ms 700 µJ laser irradiation of the dye RGA-30. 10 episodes superimposed; 80% of the pulses elicited action potentials and 100% elicited depolarizations. Depolarizing and hyperpolarizing current pulses (labelled black bands) were injected during each episode in b[ii] and b[iii]. (c)[i] Intracellular recording from an untreated Retzius cell during laser irradiation. (c)[ii] Intracellular recording from a Retzius cell with powdered carbon, but with the laser irradiation positioned to avoid illumination of any particles. Current pulses (labelled black bands) were injected after each laser stimulation (duration indicated by red dashed lines) in c[ii] and c[iii].

Important requirements for photoactivation to be useful as a tool are reliability and repeatability. By using an optical design that creates a near-diffraction limited spot of less than 2µm x 2µm x 4µm, we were able to irradiate specific carbon particle targets with high reliability in the XY plane and with reliability in Z limited by the depth of the focal plane of our imaging system. Figure 2a[ii] shows 10 intracellularly recorded episodes of photoactivation in Retzius (Rz) cells, one of which elicited an action potential and all of which elicited depolarizations. Figure 2a[iii] shows similar responses in AP cell photoactivation, in which 36/40 stimuli elicited action potentials and 40/40 elicited depolarizations. Figure 2b[ii] shows comparable experiments with RGA-30 stained Rz cells, where 2/10 stimuli elicited action potentials and 6/10 elicited depolarizations. Figure 2b[iii] shows photostimulation of an RGA-30 stained AP cell, where 8/10 stimuli elicited action potentials and all of which elicited depolarization. Current pulses were presented to the neurons to confirm normal electrophysiological behavior during each episode (Figure 2a[ii, iii], Figure 2b[ii, iii]). Powdered carbon induced photoactivation more reliably than RGA-30; with powdered carbon 88% of stimuli (12 cells, 88 trials) resulted in successful photoactivation (defined as depolarization and subsequent recovery), while in RGA-30 dye-stained neurons 65% of stimuli (12 cells, 114 trials) resulted in successful photoactivation. To verify that endogenous absorption by the cell membrane would not induce photoactivation, we measured the membrane potential during laser irradiation without any thermal absorbers present (Figure 2c[i]) and with the thermal target intentionally missed (Figure 2c[ii]); no change was observed in the membrane potential in either case, although the responses to both depolarizing and hyperpolarizing current pulses were normal.

To test the stability of photoactivation responses, we recorded cell membrane potentials of T and AP cells during a series of laser irradiations (Figure 3a[i]), each of which was immediately followed by an electrically induced depolarization (Figure 3a[ii]). Evidence of damage was typically visible as a reduction in resting membrane potential (RMP) by the tenth laser pulse (Figure 3a[ii]). As a metric of cell health, we measured the action potential amplitude change (Figure 3b[i]) and RMP (Figure 3b[ii]) after repeated irradiation. We normalized action potential amplitude and RMP to the first response to laser irradiation, then quantified the number of irradiation episodes that could be given at a particular target site before the measures fell below 75% of the first response, our criterion level for "healthy" responses. Action potential height and RMP remained above this level for 6 ± 3 episodes (6 cells, 85 stimuli) of photoactivation at 500 µJ when mediated by carbon powder. RGA-30 produced healthy action potentials and RMP for slightly fewer trials (4 ± 3, 6 cells, 44 stimuli) at the same energy. In experiments mediated by carbon particles where new targets were chosen after each pulse, successful depolarization was observed for as many as 40 pulses without permanent changes in RMP (e.g., Figure 2a[iii]).

When using powdered carbon as the thermal absorber, particle size affects depolarization magnitude: as the size of the particle increased, the amount of thermal energy absorbed increased and the depolarization became larger. For particles larger than 15 - 20 µm, energies above 500 µJ resulted in loss of membrane potential and cell death. In all cases, energies over 2000 µJ resulted in cell death. Photoactivation of RGA-30 stained neurons also produced larger depolarizations at higher laser energy levels, with cell death occurring at irradiation energies above 2000 $\mu$J. Increases in the concentration of RGA-30 during the staining process resulted in less healthy preparations, while reducing the concentration resulted in fewer laser elicited action potentials. Unstained cells were unaffected by irradiation even with pulse energies of 3500 $\mu$J.

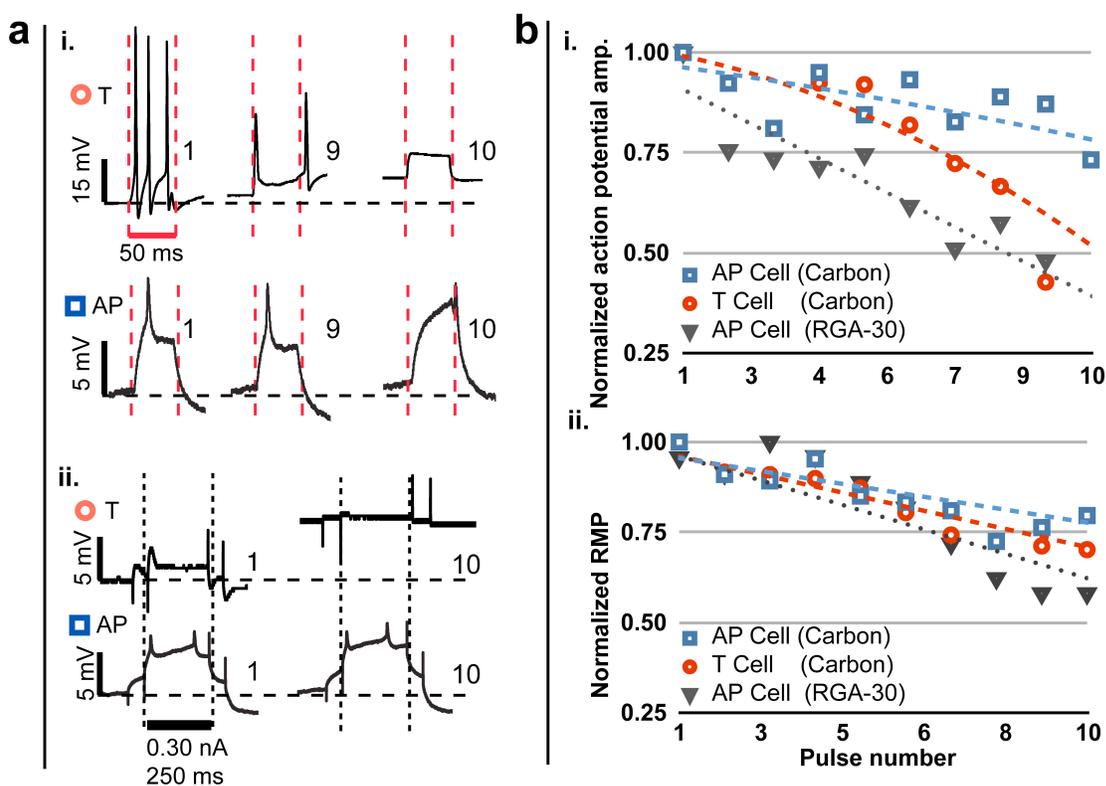

**Figure 3: Effects of repeated irradiation**
(a)[i] Photoactivated potentials induced by repeated irradiation (duration indicated by red dashed lines) of carbon particles shown during laser pulse 1, 9, and 10 in AP and T cells. (a)[ii] Electrophysiological responses to injected current pulses (vertical black lines) immediately after laser pulse 1 and 10 in AP and T cells. The initial resting membrane potential (RMP) is indicated by the dashed line. In both examples, current injection produces a healthy response after the first laser pulse. The AP cell remains healthy after laser pulse 10, while the T cell shows a decrease in response to injected current and an increase in RMP, indicating a decrease in cell health.
(b)[i] Action potential amplitude (normalized to the first observed action potential) during repeated laser irradiation. Carbon particle-mediated responses observed in the AP and T cells. Responses observed in an AP cell stained with RGA-30 are shown for comparison.
(b)[ii] RMP during repeated laser irradiation (normalized to first observed RMP). Pulses were 50 ms and 500 $\mu$J.

### B. Photoactivation occurs in the absence of voltage-gated conductances

The similarity between the responses to carbon particle-mediated photoactivation and RGA-30 stain-mediated photoactivation suggests that the production of laser-induced conductances does not depend on specific chemical in-teractions with ion channels. To distinguish between temperature-dependent activation of voltage-gated channels and conductance increases caused by other membrane changes, we measured transmembrane currents during laser exposure using two-electrode voltage clamp (TEVC) in neurons (which have voltage-gated channels) and in *Xenopus* oocytes (which have negligible levels of voltage-gated channels [18]). Irradiation of Retzius neurons elicited a small depolarizing current and an increase in membrane conductance, i.e., the current-voltage (I-V) curve of Rz cell became steeper (Figure 4a). *Xenopus* oocytes with carbon particles on the defolliculated surface and with the dark pole targeted (thus maximizing thermal absorption at the membrane) exhibit a symmetric decrease in membrane resistance proportional to the difference between the resting and holding voltage (Figure 4b). This is qualitatively similar to changes in the I-V curve of leech neurons during laser irradiation, and is strong evidence that voltage-gated channels are not necessary to produce photoactivation in response to thermal energy deposition.

## IV. Discussion

From our experiments, we conclude that the most likely cause for photoactivation of neurons is an increase in the local temperature of the membrane. Our experiments, with carbon layers under saline on a glass side, reveal that focal irradiation produces fast local temperature increases. When carbon particles placed on the surface of neurons are focally irradiated with similar levels of power, the cells depolarize and produce action potentials. Irradiation of neurons stained with RGA-30 produce qualitatively similar electrical changes, but fewer pulses can be delivered to a target cell before the membrane potential irreversibly depolarizes.

Voltage-clamp measurements of the photoactivation of both neurons and *Xenopus* oocytes mediated by powdered carbon tested whether voltage-gated channels or voltage-independent (leak type) channels are responsible for laser-induced photoactivation. Our measurements show that the effect of laser irradiation of thermal absorbers in contact with neurons or oocytes is a symmetric increase in the slope of the I-V relationship. This indicates it is unlikely that a gated or rectifying channel is responsible, as this would appear as an asymmetry in the I-V relationship.

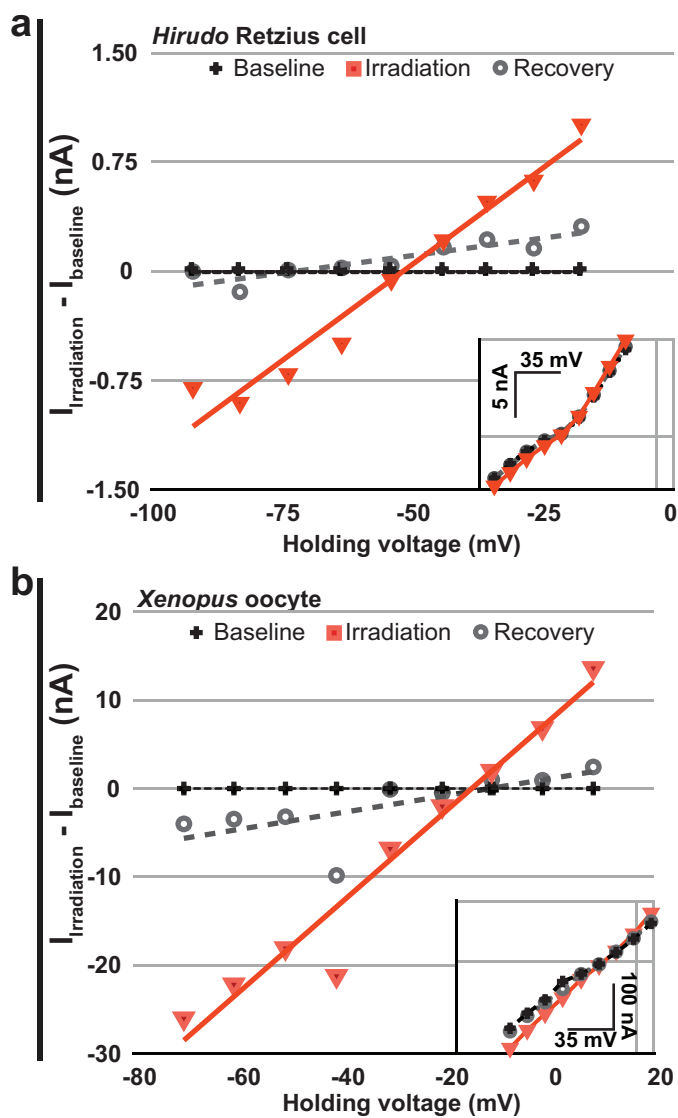

**Figure 4:**
Transmembrane currents elicited by laser photostimulation
(a) Change in transmembrane current relative to baseline elicited by a 50 ms, 250 µJ laser pulse in leech Retzius cell, measured by two-electrode voltage clamp. $R^2 = 0.96$ (irradiation). Current change after stimulation (recovery) is measured immediately after each pulse. I-V relationship shown in inset. (b) Change in transmembrane current relative to baseline elicited by a 50 ms, 250 µJ laser pulse in a Xenopus oocyte. $R^2 = 0.95$ (irradiation). I-V relationship shown in inset.

The dominant conductance in a *Xenopus* oocyte is a voltage-independent $K^+$ leak channel. To determine whether membrane changes or conductance increases in $K^+$ channels are responsible for changes in potential, we used the Goldman-Hodgkin-Katz equation and an ideal ionic gas model to predict changes to the membrane potential of an oocyte during irradiation. Assuming membrane properties remain constant and only leak conductances are present, the model predicts that the resting membrane potential would become more negative and the transmembrane electric field would be reduced as the temperature is increased. This would appear as a hyperpolarizing current during laser irradiation (a shift to the right in the I-V curve). This is contrary to our measurements in the oocyte, where we observed small depolarizing currents during irradiation (Figure 4b).

Voltage-gated channels are not temperature independent (Liang et al[19]: near-infrared low intensity laser heating in the absence of exogenous compounds increases the probability of $Na^+$ and $K^+$ channels opening. However, continuous near-infrared irradiation results in small currents that appear over tens of milliseconds. In our experiments, large depolarizations rise and decay with time constants between 1 - 3 ms. Rapid heating to equilibrium and rapid return to bath temperature would be expected for a small focal volume in contact with a large bath, such as our system, and indicates a mechanism unrelated to that observed during continuous near-infrared irradiation.

The remaining common element between leech neuronal membranes and oocytes is the lipid bilayer. Heat could have a variety of effects on the bilayer, including making transient holes in it. A more subtle effect would be on the structure of the lipids. Experiments in artificial lipid bilayers and giant unilamellar vesicles[20] reveal that some lipids have phase transitions between 20 °C and 50 °C. During these phase transitions, "rafts" of one phase can be created within an "ocean" of another. In this mixed phase, membrane conductance increases have been observed[21]. Such pores would be small, transient, and nonspecific [11], potentially allowing smaller ions such as sodium to pass through with lower resistance. This type of phase melting would result in increased non-specific conductances and a slight shift in membrane potential toward 0 mV. Although we cannot conclude that laser elicited transmembrane currents are exclusively due to phase transitions without substantial further work, our experiments strongly suggest that direct thermal interaction with voltage-dependent and voltage-independent channels are not the primary mechanism causing changes in potential.

Whatever the mechanism of the depolarizations produced by photoactivation, the crucial result is that such activation occurs in the presence of elements that are common to cells of all kinds. With an appropriate delivery system, thermal photoactivation could be implemented in many animal models. This is a complementary mechanism to stain-mediated and genetically-mediated techniques of photosensitizing neurons because it can be used without the development of genetic expression techniques. It also allows for multiple stimuli to be issued before damage occurs, and exhibits a graded response capable of sub-threshold depolarization.

Our method provides a starting point for the evaluation of thermal photostimulation methods and can be integrated with modern optical techniques. By modifying an existing design for optical tweezers, we were able to build a highly precise dual-target photoactivation system with off-the-shelf optical components. Our design is capable of constant-power stimulation of individ-ual neurons, and in single-layer preparations (such as neurons grown *in vitro*) can photoablate target cells without causing collateral damage. The design outlined could be expanded to use fully computer-controlled light modulators[2]. This would allow automated targeting of any number of neurons, limited only by aiming speed and available laser power. We believe that with reasonable costs and broad applications, thermal photoactivation has the potential to become an important technique in neuroscience.

This work was supported by the Department of Energy (MD), NIH research grant MH43396 (WBK), NIBIB T-32 Interfaces Initiative 5T32EB009380-03 (BJM), and the NSF GK-12 Program 742551 (BJM). We thank Dr. Michael Sailor, Dr. Paul Slesinger, Dr. Nicholas Spitzer, Dr. David Kleinfeld, Dr. Andrew McCulloch, and Dr. Krista Todd for their many contributions.